\documentclass[floatfix,reprint,amsmath,amssymb,prapplied,superscriptaddress,aps]{revtex4-2}

\usepackage[dvipsnames]{xcolor}
\usepackage{graphicx}
\usepackage{amsmath,amssymb}
\usepackage{dcolumn}
\usepackage[utf8]{inputenc}
\usepackage[T1]{fontenc}
\usepackage{bm}
\usepackage{fontawesome5}
\usepackage{siunitx}
\usepackage{hyperref}

\hypersetup{
colorlinks=true, breaklinks=true, 
urlcolor=RoyalBlue, linkcolor=RoyalBlue, citecolor=RoyalBlue, 
pdftitle={},
pdfauthor={\textcopyright}, 
pdfsubject={},
pdfkeywords={}, 
pdfcreator={pdfLaTeX},
pdfproducer={LaTeX}
}

\begin{document}

\title{Coercivity-size map of magnetic nanoflowers: spin disorder tunes the vortex reversal mechanism and tailors the hyperthermia sweet spot}

%% Group authors per affiliation:
\author{Elizabeth M. Jefremovas}
\email{elizabeth.jefremovas@.uni.lu}
\affiliation{Department of Physics and Materials Science, University of Luxembourg, 162A Avenue de la Faiencerie, L-1511 Luxembourg, Grand Duchy of Luxembourg}
\affiliation{Institute for Advanced Studies, University of Luxembourg, Campus Belval, L-4365 Esch-sur-Alzette, Luxembourg}
\author{Lisa Calus}
\affiliation{DyNaMat, Department of Solid State Sciences, Ghent University, 9000 Ghent, Belgium}
\author{Jonathan Leliaert}
\email{jonathan.leliaert@ugent.be}
\affiliation{DyNaMat, Department of Solid State Sciences, Ghent University, 9000 Ghent, Belgium}

%%%%%%%%%%%%%%%%%%%%%
\begin{abstract}
Iron-oxide nanoflowers (NFs) are one of the most efficient nanoheaters for magnetic hyperthermia therapy (MHT). However, the physics underlying the spin texture of disordered iron-oxide nanoparticles beyond the single-domain limit remains still poorly understood. Using large-scale micromagnetic simulations we completely map the magnetization of NFs over an unprecedented size range, from 10 to 400 nm in diameter, connecting their microstructure to their macroscopic magnetic response. Above the single domain ($d >$ 50 nm), the magnetization folds into a vortex state, within which the coercivity describes a secondary maximum, not present for non-disordered nanoparticles. We have extended our understanding by resolving also the NF magnetization dynamics, capturing the physics of the magnetization reversal. Within the vortex regime, two distinct reversal modes exist: i) A core-dominated one, in which the core immediately switches along the direction of the applied field, resulting in an increasing coercivity for larger sizes; and ii) a flux-closure dominated reversal mode, going through the perpendicular alignment of the vortex core to the field, resulting in a decreasing coercivity-size dependence. The coercivity maximum is located at the transition between both reversal modes, and results from the combination of grain anisotropy and grain-boundary pinning: weak (but non-negligible) inter-grain exchange keeps the vortex profile coherent, yet allows the core to be pinned by the random anisotropy easy axes of the single grains, maximizing magnetic losses. Our results provide the first full description of spin textures in iron oxide NFs beyond the macrospin framework, and clarify the role of internal spin disorder in magnetic hyperthermia heating. By adjusting the grain size, the coercivity "sweet spot" can be tailored, offering a practical route to next-generation, high-efficiency nanoheaters.
\end{abstract}

\maketitle

%%%%%%%%%%%%%%%%%%%%%%%%%%%%%%%%%%%%%%%%%%%%%%%%%%%%%%%%%%%%%%%%%%%%%%%%%%
\section{Introduction}

Magnetic nanoparticles (MNPs) are at the forefront of a new era in personalized medicine, hosting potential in both diagnostic and  therapeutic applications~\cite{wu2019magnetic, etemadi2020magnetic, farzin2020magnetic, lapusan2024advancing, rezaei2024magnetic,coene2022magnetic}. Their highly tunable magnetic response and size enables targeted approaches, such as magnetic particle hyperthermia (MHT), where local heating selectively disrupts malignant cells, providing a promising complement to conventional cancer therapies~\cite{rubia2021whither, issels2008hyperthermia, diaz2025preclinical, durando2024combination, wu2024roadmap}. Among MNPs, iron oxide nanoparticles (IONPs) are especially well suited owing to their excellent biocompatibility and magnetic tunability, which together enable the design of personalized treatment strategies aligned with the requirements of next-generation nanomedicine~\cite{anselmo2016nanoparticles, anselmo2019nanoparticles, anselmo2021nanoparticles, soetaert2020cancer, berns2020towards}.

A key challenge in MHT is to maximize the heating power of IONPs, achieving effective thermal responses with minimal nanoparticle dosage. 
Several strategies have been explored to achieve this aim, such as tuning the composition, size, shape, and degree of disorder, as well as engineering supra-particle arrangements like chains \cite{roca2019design, gavilan2021magnetic, jefremovas2021nanoflowers, gandia2019unlocking, mekseriwattana2025shape, coene2022magnetic, lappas2019vacancy}. 
Nonetheless, these efforts have often been carried out without a full understanding at microscopic level of how these factors influence magnetic heating performance, thereby limiting their predictive power.
In particular, both the spin texture, i.e., the spatial arrangement of magnetic moments inside the nanoparticle, and the dynamic response of that texture to externally applied magnetic fields are essential for heat dissipation, yet their roles remain poorly understood. Bridging the nanoparticle microstructure to its macroscopic magnetic properties, such as coercivity, is the bottle-neck for the rational design of next-generation hyperthermia agents.

One particular spin texture highly relevant for biomedical applications is the vortex state. This inhomogeneous spin texture minimizes stray magnetic fields, suppressing interparticle dipolar interactions, thereby preventing undesired magnetic aggregation~\cite{serantes2021nanoparticle, etheridge2014accounting}. At the same time, the vortex core retains a net magnetization, which can contribute to heating or serve as a vector for drug delivery~\cite{ho2011monodisperse, goiriena2016high, goiriena2020disk, usov2018magnetic}. Theoretically, vortex states are predicted in fine ferromagnetic nanoparticles exceeding the single-domain size limit~\cite{di2012generalization, tauxe2002physical, brown1969fundamental, brown1978domains}, with the critical size depending on material properties~\cite{schabes1988magnetization, hertel2002finite, rave1998magnetic, magnetosomes2013, betto2014vortex, gan2014multi, kakay2005monodomain, witt2005three}. Experimentally, vortex states have been observed in IONPs of different morphologies, including disks, ellipsoids, spheres or cubes~\cite{usov2018magnetic, gao2020ellipsoidal, lewis2020magnetic, goiriena2020disk, concas2024magnetic}.
More recently, vortex states were detected in multicore nanostructures called nanoflowers (NFs)\cite{moya2024unveiling}. NFs perform exceptionally in MHT~\cite{jefremovas2021nanoflowers, bender2018relating, vivas2020toward, hugounenq2012iron}, and their synthesis is already scaled to commercial production~\cite{storozhuk2021stable, gavilán2017formation, mekseriwattana2025shape}, making them strong candidates for clinical translation in magnetic hyperthermia~\cite{prajapati2024transforming}.

Despite growing interest in vortex configurations in nanoparticles, the theoretical interpretation of experimental observations remains incomplete. Detecting clearly and unambiguously which features the models should capture to accurately explain the magnetization in complex morphologies, like nanoflowers, remains challenging. Most theoretical studies employ idealized models that include only demagnetizing and exchange interactions (e.g. Ref.~\cite{di2012generalization}). Experiments, however, reveal that anisotropy and spin disorder, both intrinsic to nanostructured systems, play key roles in stabilizing non-uniform spin textures~\cite{vivas2020toward, gatel2015size, pratami2023micromagnetic, lak2021embracing}. More precisely, spin disorder, originating at magnetic imperfections, and always present in nanoparticles as a consequence of surface effects (lower symmetry, reduced coordination), promotes inhomogeneous magnetic patterns, yet its influence on magnetization dynamics remain unclear~\cite{lima2006spin, khurshid2012surface, lak2021embracing, zakutna2020field}. Elucidating the physics of intra-particle spin disorder is crucial to exploit its full application potential, as theory predicts that magnetic defects could raise the heating efficiency by up to an order of magnitude compared with defect-free nanoparticles~\cite{lappas2019vacancy}.

In this work, we go beyond the macrospin framework and perform a full micromagnetic numerical study to resolve the inhomogeneous spin texture of magnetic nanoflowers, revealing how intra-particle spin disorder governs their magnetization reversal. By connecting local magnetic inhomogeneities to macroscopic coercivity, we demonstrate that tuning the degree of internal disorder enables control over the threshold NF size required to stabilize vortex states, as well as the NF size at which coercivity is maximized. %Our study uses the magnetic parameter of coercivity, $\mu_{0}H_{C}$, as a proxy for the energy stored as the magnetization goes through the hysteresis loop.  
Our findings bridge the nanoparticle microstructure, determined by the spin texture, and heating performance, represented by the coercivity, offering unique and fundamental insights to tailor the coercivity “sweet spot” for optimized hyperthermia performance.

\begin{sloppypar}
The paper is structured as follows: in Section \ref{primera} we present the full coercivity vs. NF diameter map, from diameter $d=10~\mathrm{nm}$  to $d=400~\mathrm{nm}$, identifying the secondary maximum of coercivity for the vortex regime. Section \ref{segunda} details the magnetization reversal mechanisms of the vortex phase, revealing two distinct switching regimes determined by NF size. Finally, Section \ref{tercera} elucidates how the grain size and other disorder details like the exchange coupling reduction at grain boundaries affect coercivity.
 \newline
\end{sloppypar}

%%%%%%%%%%%%%%%%%%%%%%%%%%%%%%%%%%%%%%%%%%%%%%%%%%%%%%%%%%%%%%%%%%%%%%%%
\section{Results and Discussion}

Our numerical study is designed to mirror the structure and behaviour of maghemite nanoflowers as realistically as possible. From a structural perspective, nanoflowers present a 
nearly-spherical shape, which preserves the advantages of single-core spherical nanoparticles: almost isotropic coercivity and uniform heat release irrespective of the field orientation~\cite{poon2025cubic}. Moreover, the rough surface and dense network of grain boundaries creates an intraparticle energy landscape rich in pinning sites, allowing spin disorder to influence the magnetization dynamics. We have included in Figure~\ref{fig_1}\textbf{A} as an inset a representative image of our simulations, which reproduce such unique and distinct multicore morphology. As such, we have modeled the NF as a quasi-spherical cluster of crystalline grains, each with its own uniaxial anisotropy axis $K_{u}$ (indicated by different colours). We have turned on the parameter of spin disorder by lowering the exchange stiffness $A$ at the grain boundaries with a scaling factor $k$ ($0 \le k \le 1$), so the intra-grain stiffness is $A$, and inter-grain stiffness becomes $kA$. This parameter lets us systematically probe how pinning affects the magnetization dynamics and, ultimately, heating performance.

%Accordingly, our study resolves the spin texture of maghemite NFs, with the purpose of connecting their particular microstructure, accounting for spin disorder, to their potential heat release. For this, we base our study on the magnetic parameter of coercivity, $\mu_{0}H_{C}$, a proxy for the energy stored as the magnetization goes through the hysteresis loop. We begin our study by following $\mu_{0}H_{C}$ for a collection of NF sizes ranging from 10 to 400 nm, which allows to determine the size regime relevant for therapeutical applications, where the coercivity reaches a maximum. We then focus on this window and resolve the switching mechanism to connect the role of intra-particle spin disorder and the maximized energy losses. Finally, we modulate intra-particle disorder to determine its impact to the coercivity maximum.

%%%%%%%%%%%%%%%%%%%%%%%%%%%%%%%%%%%%%%%%%%%%%%%%%
\subsection{Coercivity vs. NF size}\label{primera}

\begin{figure*}
\centering
\includegraphics[width=0.8\linewidth]{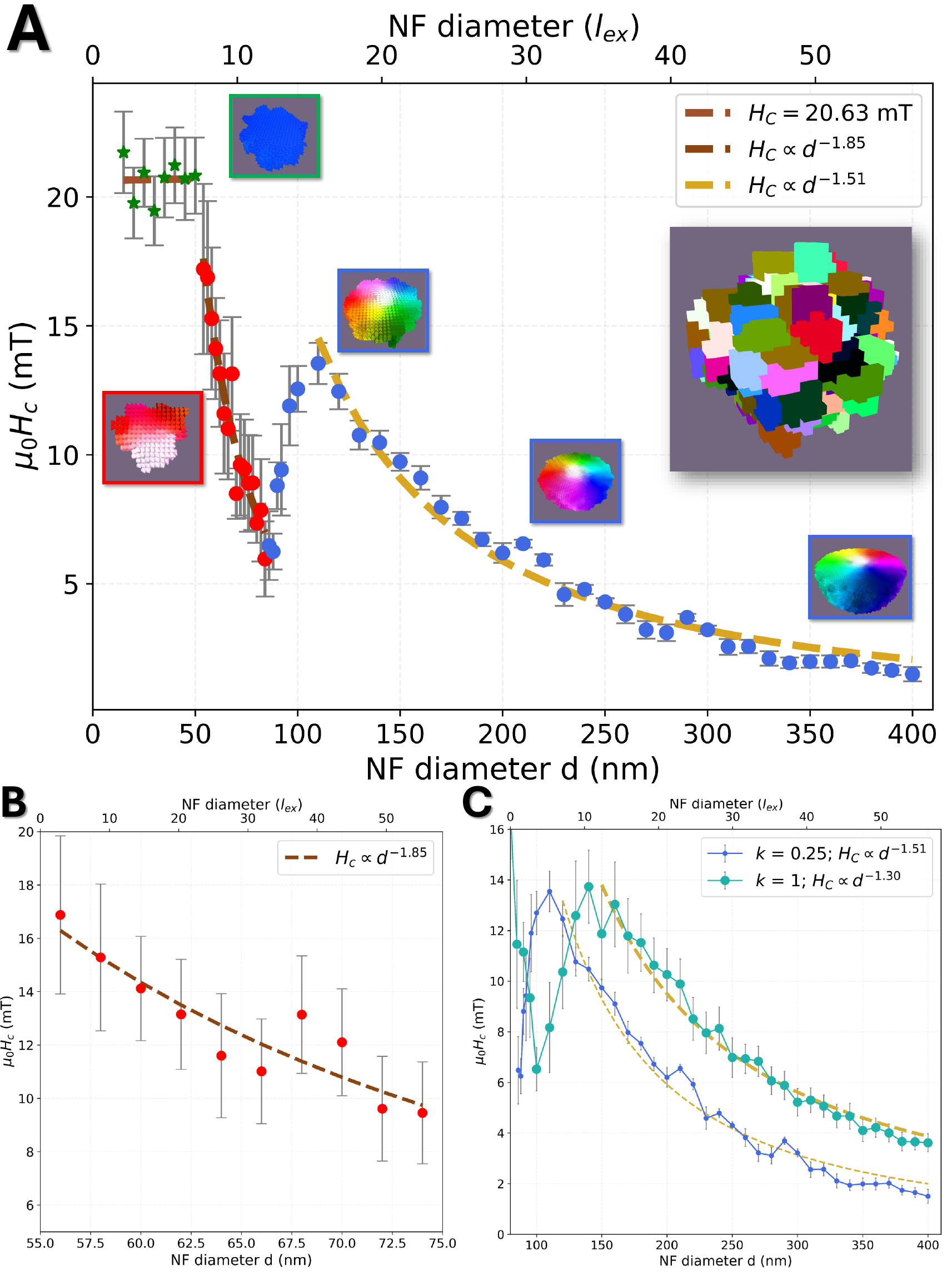}
\sloppy
  \caption{\textbf{(A)} Coercivity map as a function of NF diameter for $k$=0.25 and grain size = 15 nm. Beyond the single-domain Stoner-Wohlfarth regime (green stars, $d<50~\mathrm{nm}$ nm, single grain) there appears another single-domain regime, but with a reversal not via coherent rotation (red dots, $50~\mathrm{nm}<d<75~\mathrm{nm}$). For even larger diameters, a vortex is stabilized within the whole range (blue dots). The insets show the micromagnetic structure (equatorial plane) at remanence ($\mu_{0}H_{0}=~0~\mathrm{T}$) for $d = 30, 70, 120, 250 ~\mathrm{and}~ 400 ~\mathrm{nm} $. After reaching a maximum at $d=110~\mathrm{nm}$, the coercivity drops $\propto d^{-1.5}$. Top right inset includes the simulated grain microstructure of NF $d=100~\mathrm{nm}$, where the uniaxial anisotropy direction for each grain is represented in different colors. \textbf{B} zooms-in on the single-domain to vortex region, where the coercivity drops $\propto d^{-1.85}$ before the rise in the vortex regime. \textbf{C} Comparison between the $k=0.25$ and $k =1$ cases. The onset of the vortex, together with the maximum of the coercivity (and thus, of the hyperthermia sweet spot), is shifted to larger values at $k = 1$.}
  \label{fig_1}
\end{figure*}

Figure \ref{fig_1} maps the evolution of the coercive field, $\mu_{0}H_{C}$, as a function of the nanoflower diameter $d$. Diameters are plotted both in absolute units (nm, bottom axis) and normalized to the exchange length, $l_{ex}=\sqrt{2A/\mu_{0}M_{s}^{2}}$ (top axis), allowing material-independent comparison and bridging to intrinsic magnetic length scales.

Unless otherwise specified, all simulations use a grain size of 15 nm and an inter-grain coupling factor $k=0.25$, yielding excellent agreement with experimental data~\cite{jefremovas2021nanoflowers, moya2024unveiling}. For comparison, a fully coupled case ($k=1$) is also included to isolate the effect of grain boundaries while preserving the nanoflower geometry. Diameters from 10 nm to 400 nm are examined, fully covering the relevant experimental size range~\cite{jefremovas2021nanoflowers, gavilan2021magnetic, gavilán2017formation, gavilan2021size, moya2024unveiling, bender2018relating, hugounenq2012iron, borchers2025magnetic}. Representative complete hysteresis loops for 20 different random initial configurations per NF size are included in the Supplemental Material S2.

\begin{sloppypar}
Figure \ref{fig_1}\textbf{A} reveals three size regimes, each with a distinct, qualitatively different, coercivity trend.
In the smallest regime ($d<$ 50 nm, green stars) the diameter lies below the single domain limit. Thereby, we model every nanoflower as a single grain. Simulations that explicitly resolve 15 nm grains are methodologically challenging because of the small and irregular number of grains per particle. To confirm that our approximation is adequate, we have included the results corresponding to the simulations performed with such grain size at this low $d$ regime in the Supporting Information, section S3. There, Figure~S2 shows that even at the upper bound ($d=50$ nm, $\approx 15$ grains) the grain structure does not result in any deviations from the single domain and single grain behaviour.

In our single grain simulations, the magnetization remains uniform throughout the entire hysteresis loop (see snapshot included in green at remanence), indicating that the reversal mechanism is a coherent rotation. Under these conditions the data can be interpreted using the Stoner-Wolhfarth (SW) model \cite{tannous2008stoner}, where the coercive field remains constant and is given by $\mu_{0}H_{C}= 0.48 \frac{2K}{M_{s}}$, being the factor $0.48$ included to account for the random distribution of the anisotropy axes\cite{carrey2011simple}. For our material parameters, this idealized model predicts a constant $\mu_{0}H_{C} = 24~\mathrm{mT}$, which is slightly larger but in reasonable agreement with the numerical results of $H_{C}=20.6~\mathrm{mT}$.
\end{sloppypar}

The second regime spans between $50~\mathrm{nm} < d <75~\mathrm{nm}$ (red dots in Figure~\ref{fig_1}\textbf{(A)}). Once the diameter exceeds the single-domain threshold of $7.2 l_{ex}$ \cite{di2012generalization} (51 nm for our material) reversals do not longer happen through a coherent rotation, but rather through domain wall nucleation (see inset marked in red), which lowers the coercive field relative to the Stoner–Wohlfarth limit observed for smaller particles. Unlike the constant value of the coercivity within the SW regime, at this regime, the coerivity scales with size as $H_{C}\propto d^{-1.85}$. This change in the coercivity dependence is a direct consequence of the nanoflower size exceeding the single-domain regime. Accordingly, above $d =50~\mathrm{nm}$, the internal grain structure of the NFs, with each grain having an independent random anisotropy easy axis, becomes dominant. The random anisotropy axes average the total anisotropy out to an effective anisotropy, $K_\text{eff} \propto 1/ \sqrt{N}$, where $N$ is the number of grains. Because a quasi-spherical nanoflower contains $N \propto V \propto d^3$ grains, $H_{C}\propto K_\text{eff} \propto d^{-\frac{3}{2}}$. The fitting of our simulated results, included in Figure~\ref{fig_1}\textbf{(B)}, yields $H_{C}\propto d^{-1.85}$, indicating a reasonable agreement with this simplified model. We attribute the slightly steeper decline than the -1.5 power law expected for perfect spheres to disorder associated with the irregular edges of the nanoflowers.  \newline

The third regime (blue symbols in Figure~\ref{fig_1}\textbf{A}) starts once the nanoflower reaches a diameter large enough to host a vortex state at remanence (see blue insets). In our material, this happens for $d_{\text{vortex}}\gtrsim 70~\mathrm{nm}$ ($\approx 9.9\,l_{ex}$), consistent with previous reports~\cite{schabes1988magnetization, hertel2002finite, rave1998magnetic, magnetosomes2013, betto2014vortex, gan2014multi, kakay2005monodomain, witt2005three}. This threshold exceeds the $d \ge 7.2\,l_{ex}$ predicted for ideal, isotropic spheres~\cite{di2012generalization}, underscoring the influence of crystalline anisotropy. While vortex formation lowers the demagnetizing energy, magnetic anisotropy imposes an energy penalty. The balance shifts in favor of the vortex only when the particle is sufficiently large for the demagnetizing energy to dominate, hence the larger critical size.

To isolate the anisotropy effects from those connected to exchange within this regime, we simulated two reference systems: a nanoflower (NF) which preserves the realistic irregular morphology, but whose grains are fully exchange-coupled ($k=1$); and additionally, a perfect sphere geometry with uniform uniaxial anisotropy $K_{u} \parallel z$, referred to as NS. The former case (fully-coupled NF) is shown in Figure~\ref{fig_1} \textbf{C}, allowing a fair comparison to the case with inter-grain boundaries ($k =$ 0.25), while the NS case is discussed in detail in the Supplemental Material S5.

In both these reference systems, the vortex nucleates at larger sizes compared to our standard $k =0.25$ case, being $d_{vortex}>100$ nm for $k =1$ and $d_{vortex}>93$ nm for NS. The reduced size for a vortex to nucleate at $k = 0.25$ follows directly from the scaling of the critical diameter with the exchange length, $d_{vortex}\propto l_{ex}=\sqrt{2A/\mu_{0}M_{s}^2}.$ The lowering of the intergrain exchange $kA$ decreases the effective exchange coupling, shortens $l_{ex}$, and therefore decreases $d_{vortex}$. In other words, weakening the exchange coupling at the grain boundaries reduces the energetic cost of non-uniform spin textures, enabling the stabilization of the vortex state in smaller nanoflowers.

%Within the vortex regime, we identify two distinct trends in the coercivity as a function of the NF diameter. Immediately after the vortex stabilization, the coercivity increases sharply, as a consequence of the finite net magnetization aligned along the $z$-axis at the vortex core. This core magnetization imposes an additional demagnetizing energy penalty, thereby raising the coercive field. Notably, this rise is absent in the case of a perfect sphere (NS), where the magnetization drops continuously once the nanoparticle exceeds the single domain regime (see Supplemental Material \textcolor{green}{Supplemental}). This finding reveals the critical role of structural disorder behind the hyperthermia "sweet spot". Even in the absence of explicit intra-particle grain boundaries ($k =$ 1), the presence of grains with randomly oriented anisotropy axes acts as effective pinning sites for the vortex core magnetization. These introduce internal energy barriers resulting in a non-zero coercivity and thus, enhancing magnetic losses. 

Within the vortex regime, two distinct coercivity trends emerge, in both $k =0.25$ and $k = 1$, as the nanoflower diameter increases. Just above the vortex nucleation threshold the coercivity rises sharply because the magnetization becomes more vortex-like (less domain-wall-like) with size. As elaborated on in Section~\ref{segunda}, the sharper features of the vortex core are pinned more easily by material disorder. The larger NF diameter implies an increase of the number of constituent grains and, hence, the number of random anisotropy directions, which act as pinning sites for the vortex core, resulting in the raise of the coercivity. The growth continues until a maximum, the ``hyperthermia sweet spot'' is reached at $d=110$ nm for $k=0.25$ and $d=150$ nm for $k=1$. Note that in both cases, the coercivity peaks at the same values ($\cong 14~\mathrm{mT}$), endorsing the hypothesis that the main factor behind this raise is the existence of grains within the material. Even when every grain is fully exchange-coupled ($k=1$), the randomly oriented easy axes provide pinning potentials for the vortex core, introducing internal energy barriers resulting in a non-zero coercivity and enhanced magnetic losses. Our hypothesis is further supported when comparing to the case of a perfect sphere (NS), where the raise is absent and the coercivity drops continuously once the nanoparticle exceeds the single domain regime (see Supplemental Material S5). This contrast underscores how the hyperthermia sweet spot critically depends on structural disorder. 

Beyond the "sweet spot diameter", the coercivity decreases, following a power-law decay, scaling as $d^{-1.5}$ for $k=0.25$ and $d^{-1.3}$ for $k=1$, in agreement with previously reported scalings for the vortex regime~\cite{noh2012nanoscale, sung2015magnetic, demortiere2011size, tauxe2002physical, kneller1963particle, hergt2008effects, heider1987magnetic}. This non-monotonic behaviour marks a transition in the dominant magnetization reversal mechanism, which we analyse in detail in the next section.

%%%%%%%%%%%%%%%%%%%%%%%%%%%%%%%%%%%%%%%%%%%%%%%%%%%%%%%%%%%%%%%%%%%%

\subsection{Magnetization reversal in the vortex regime}\label{segunda}
%In Section~\ref{primera}, we resolved the evolution of coercivity as a function of nanoflower (NF) size, revealing a distinct twofold behavior (an initial rise followed by a drop) within the vortex regime. Comparison with the case of ideal spherical particles (NS) revealed two separate scaling behaviours in the decay of coercivity, suggesting a change in the underlying magnetization dynamics. To elucidate the origin of such dual behavior, we now analyze the magnetization reversal mechanisms across different NF sizes, aiming to identify size-dependent effects.

In Section~\ref{primera} we investigated the coercivity as a function of nanoflower (NF) size, finding a non-monotonic trend within the vortex regime: an initial rise up to a critical $d_{vortex}$ size, followed by a decrease. To elucidate the origin of this dual behaviour, we here examine the magnetization reversal mechanism within both of these size ranges.

Figures~\ref{fig_2} \textbf{A} and \textbf{B} depict the polar angle $\theta$ between the net NF magnetization (see Supplemental Material S4 for technical details) and the applied field direction $+z$ for nanoflowers with $d = 100~\mathrm{nm}$ \textbf{(A)} and $d = 120~\mathrm{nm}$ \textbf{(B)}. These diameters sit just below and above the critical size $d_{vortex} = 110~\mathrm{nm}$ where the vortex coercivity peaks for $k = 0.25$.

In the range $70 < d < 110~\mathrm{nm}$, where $\mu_{0}H_{C}$ is still rising, the magnetization reversal process typically\footnote{Because we always investigate several realization of randomly generated geometries, both processes happen in an overlapping size range, and we identify the peak with the transition of which process dominates the resulting trend in the coercivity.} happens in one step from the positive $+z$ to the negative $-z$ direction (yellow-marked snapshots), and spends only brief moments in intermediate states where the net moment lies perpendicular to the field (purple-marked snapshots). For an ideal sphere the switch is sharper still, showing an almost instantaneous reversal at the switching field (see Supplemental Material S5).

In contrast, when the NF size exceeds $d =110~\mathrm{nm}$, the reversal process happens gradually. Figure~\ref{fig_2} \textbf{B} shows how the vortex core rotates into the plane and remains nearly perpendicular to the applied field (see snapshot in purple) over a wide field interval, from $\approx -5~\text{mT}$ to $\approx -25~\text{mT}$ before finally flipping to $-z$. Both Figures~\ref{fig_2} \textbf{A} and \textbf{B} show that the two magnetization reversal processes are found regardless the $k$ value. Within each process, there is also not a strong $k$ dependence, as the $\theta$ \textit{vs.} $\mu_{0}H_{z}$ curves for $k =$ 0.25, 0.75 and 1 almost coincide. 

%We have illustrated the switching mechanism by including in Figure~\ref{fig_2} \textbf{C} three visualizations, from $+z$ to $-z$ through a direction perpendicular to $z$, for the $d = $110 nm, as representative of the process. The net magnetization (i.e., vortex core) is marked in red, and the flux-closure moments in gray. This switching applies for all the NFs, regardless their intra-particle pinning. 

Figure~\ref{fig_2} \textbf{C} illustrates the mechanism for $d = 120~\text{nm}$ with three representative snapshots showing the vortex core (along the net magnetization direction, shown in red, surrounded by flux-closure moments shown in gray) rotating from $+z$ through an in-plane orientation to $-z$. This sequence is identical for all nanoflowers in this regime, irrespective of intra-particle pinning.

According to our findings, we can conclude that particle size, not internal disorder, dominates the reversal mechanism. To verify this, we have simulated an ideal sphere with a single uniaxial anisotropy axis $K_{u} \parallel z$, i.e., without internal disorder. The result of these simulations are shown in Figures~\ref{fig_2} \textbf{(D)} and \textbf{(E)}, where we plot the remanent magnetization component along the field direction, $m_{z}$ (\textbf{D}) and the relative vortex core volume $V_{core}/V_{NS}$ (\textbf{E}), as a function of the NS diameter. Representative remanent states are shown between the panels.

Beyond the single domain range (marked in blue in both figures), the magnetization folds into a vortex state at ($d>d_{vortex} = 93~\mathrm{nm}$), revealing the threshold size for which the demagnetizing energy dominates exchange and anisotropy. Once the vortex is stabilized, we distinguish two different regimes, up to $d = 107~\mathrm{nm}$, and above $d = 108~\mathrm{nm}$. 

Within the first regime ($93<d<108~\mathrm{nm}$), the vortex core points along $z$ at remanence (yellow inset). Both the remanent $m_{z}$ and $V_{core}/V_{NS}$ decrease $\propto d^{-4.63}$ (red line in \textbf{E}), which is steeper than the $\propto d^{3}$ rise of $V_{NS}$. This indicates that the core volume $V_{core}$ itself is \emph{shrinking} with increasing size, as $d^{-4.63}$ is steeper than the $d^{-3}$ volumetric factor. Note that, for the NS, this shrinking core translates into a lower coercivity (see Supporting Information S5), whereas it \emph{increases} the coercivity for NFs (see Fig.~\ref{fig_1}). This opposite size dependence can be understood by considering the different anisotropy landscapes in NS and NF. In the perfect NS, the anisotropy is uniform with its easy axis parallel to the $z$ axis. The vortex core is stabilized along this direction by an energy proportional to the anisotropy constant $K$ and its volume $V_\text{core}$. As $d$ increases, the core volume shrinks, and consequently, the anisotropy barrier falls. The Zeeman energy decreases too, because only the shrinking fraction of spins along the $z$ axis can switch from an energetically unfavourable to a favourable orientation. As a consequence, the anisotropy barrier drops even faster than the volumetric $d^{-3}$, since the smaller the core, the larger the fraction of spins oriented in-plane, which do not contribute to the anisotropy. This large reduction in anisotropy energy barrier therefore explains the decrease in coercivity for NS.

%which is highly costly in anisotropy energy. 
 %d^-4.6 = d ^0.21
The anisotropy landscape in NFs is very different, as each grain has its own randomly oriented anisotropy easy axis. Thereby, the magnetization can adjust locally, more or less aligning with the anisotropy of each grain almost irrespective of the vortex core’s size within the total vortex volume. This means that in NF, there is almost no reduction in the anisotropy energy barrier with size, as the magnetization will always find a local direction to align with. As $d$ grows, and $V_\text{core}$ shrinks, but the number of grains (and thus, random directions) increases, the external field exerts a torque on the magnetization to overcome the switching energy barrier that is effectively decreasing in strength. As a consequence, the coercivity raises.

Note that pinning at grain boundaries, whose strength scales with $k$, has no qualitative effect on the coercivity in this regime: the increase appears for both $k = 0.25$ and $k = 1$, reaching the same heights of the peak in the coercive field. Thus, the rise in coercivity is driven mainly by the interplay between anisotropy effects and the vortex core size within the vortex profile. 
The latter does depend on $k$, leading to the shift in the location of the coercivity peak observed in Fig. \ref{fig_1} \textbf{C}. We defer the detailed discussion of the influence of the grain boundaries to Section~\ref{tercera}, and instead continue with the second vortex reversal regime of the ideal NS case.

The second regime starts as soon as the magnetization reversal is no longer core-dominated, which happens when $V_{core}/V_{NS}<\frac{1}{3}$. As observed in Figure~\ref{fig_2}\textbf{(E)}, this happens for diameters $d>107~\text{nm}$, when the vortex core is no longer the main volume and the flux-closure moments along the $x$ and $y$ directions carry more volume. The NS thus minimizes its energy by rotating the magnetization and aligning about half (a bit more due to deformation of the profile) of these flux-closure spins with the field ($z$) direction, while the core turns perpendicular to $z$ (see purple insets and the central snapshot in Figure~\ref{fig_2}\textbf{(C)}). As a consequence, the remanent magnetization and coercivity drop to zero.

The deformation of the vortex profile also explains the gaps between the blue, yellow, and purple regions in Figures~\ref{fig_2}\textbf{(D)} and \textbf{(E)}. While the core is aligned with the easy axis, it enlarges slightly as the gain in anisotropy energy outweighs the cost in demagnetizing energy, so this regime extends to larger $d$ than it would in the absence of uniform anisotropy $\parallel z$. After the core rotates perpendicular to the axis, its size is slightly suppressed, and also the volume in the different flux-closure directions is no longer equivalent: the volume parallel to the easy axis grows at the expense of the energetically less favorable one.

In the NF, we find the same $V_{core}/V_{NS}=\frac{1}{3}$ threshold at which the switching mechanism shifts from one in which the core immediately reverses along the $z$ direction to a rotation in which the core remains perpendicular to the $z$ axis for a considerable field range. Whereas in a NS with uniform $K_{u}$ this leads to zero net magnetization and hence vanishing coercivity, in NFs, however, the material grains pin the magnetization and postpone the coercivity drop, even though the reversal mode has changed. Within this second regime, we find $V_{core}/V_{NS}\propto d^{-1.4}$, consistent with the power law observed in Figure~\ref{fig_1} for the decrease in coercivity beyond its maximum. The optimal balance between anisotropy and rotation is therefore reached at $V_{core}/V_{NF}=\frac{1}{3}$, where equal fractions of magnetization are aligned parallel and perpendicular (2 directions) to the easy axes.

\begin{sloppypar}
We conclude that the magnetization reversal mechanism is governed primarily by the size dependence of the vortex profile: when the vortex core dominates, switching proceeds via an immediate reversal; when the flux‐closure volume dominates, reversal occurs by reorienting the flux‐closure moments along the field and anisotropy axis ($z$), and the core perpendicular to it. The threshold for both mechanisms is $V_{core}/V_{NF}=\frac{1}{3}$, which determines which part of the nanoparticle encompasses more magnetic moments. We have also found that introducing intra‐particle disorder (grains with random local easy‐axis orientations) increases the effective anisotropy, leading to superior coercivity. 
\end{sloppypar}

\begin{figure*}
  \includegraphics[width=\linewidth]{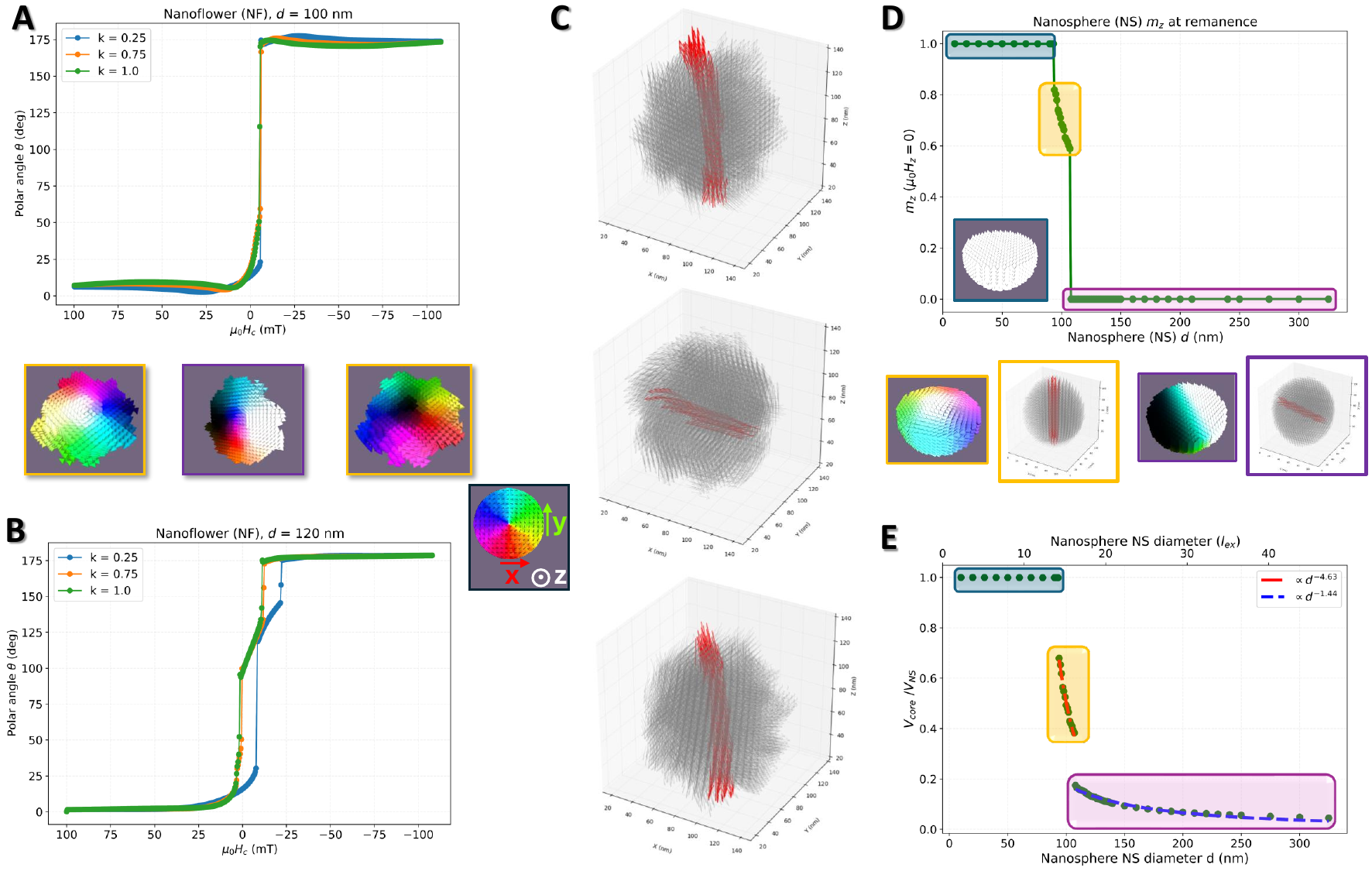}
  \sloppy
  \caption{Elucidating the magnetization reversal mechanism of the vortex state. \textbf{A} and \textbf{B} depict the polar angle $\theta$ of the vortex core direction \textit{vs.} applied magnetic field $\mu_{0}H_{z}$ for \textbf{A} at $k$ values of 0.25,0.75 and 1.0, for \textbf{A} $d= 100~\mathrm{nm}$ and \textbf{B} $d = 120~\mathrm{nm}$ nanoflowers. The snapshots represent the magnetization states: the left and right ones (with yellow border) show the vortex core along the positive (white core region) and negative (black core region) $z$ direction, respectively, whereas the middle one with purple border has the core oriented in an in-plane direction. The color wheel is included to help with the interpretation of these directions. In \textbf{(C)} we show the rotation of the vortex core for $d = 110~\mathrm{nm}$. The core direction is marked in red, and the flux-closure cells in gray. \textbf{D} and \textbf{E} show results obtained on the non-disordered nanospheres NS. \textbf{D} shows the magnetization along $z$ at remanence, and \textbf{E}, the $V_{core}/V_{NS}$ ratio as a function of the NS diameter. 
  The blue-shaded area denotes the single-domain state. In the yellow region, the vortex core aligns with the $z$-axis at remanence, whereas in the purple region, the core lies within the $XY$ plane.}
  \label{fig_2}
\end{figure*}

%%%%%%%%%%%%%%%%%%%%%%%%%%%%%%%%%%%%%%%%%%%%%%%%%%%%%%%%%%%%%%%%

\subsection{Influence of intra-particle disorder}\label{tercera}
\begin{sloppypar}
In Section~\ref{segunda}, we established that the magnetization reversal mechanism in both NF and NS vortex states is fundamentally dictated by geometric factors; essentially, the size dependence of the vortex core profile. This does not prevent however intra-particle disorder, like grain-boundaries, to play an essential role in the magnetic behaviour of the nanoparticles. Already in Section~\ref{primera}, we clearly detected that intra-particle disorder effectively shifts the $d_{vortex}$ threshold. In Section~\ref{segunda}, we observed how the difference in the anisotropy energy landscape between a perfect sphere with one easy axis and a nanoflower consisting of a large number of randomly oriented grains introduces a peak in the coercivity in the latter, whereas it monotonically drops in the former. By combining both effects, intra-particle disorder can be exploited to tune the coercivity-size relationship, and thus, the ``hyperthermia sweet spot''.

%Our earlier result issued from Figure~\ref{fig_1} \textbf{(C)} showed that varying the inter-grain exchange coupling does not lead to such qualitative differences, but does allow one to shift the onset of the vortex regime and the peak in the coercivity to different sizes.

%Specifically, the intrinsically larger anisotropy of nanoflowers, consequence of their irregular morphology and their grain texture, shifts the single‐domain to vortex transition, and accordingly, the size at which the coercivity is maximum, to larger diameters. This shifts occurs because the anisotropy constant must be sufficiently reduced (with $K \propto 1/\sqrt{N}$) so as the demagnetization energy can stabilize a vortex texture. This anisotropy shift can be modulated by reducing the inter-grain exchange coupling (and thus, incrementing the pinning). By reducing exchange coupling via $k \times A$, a weaker effective anisotropy is established, leading to an enhanced spin canting at the boundaries. This facilitates the formation of a vortex at smaller sizes, while still producing a comparable coercivity to that of perfectly coupled grains as a consequence of the pinning of the magnetization at the grain boundaries.
\end{sloppypar}

While $k$ itself is a microscopic parameter governed by atomic-scale interactions, and thus, practically not possible to engineer experimentally, we propose to control the crystallite size during the synthesis as a practical route for tuning the disorder and consequently the position of the hyperthermia peak  \cite{concas2024magnetic, attanayake2024superparamagnetic, moya2024unveiling, mekseriwattana2025shape}. By adjusting the growing conditions, it is possible to vary the size and thus the number of constituent grains $N$, which will determine the density of grain boundaries. This in turn alters the magnetic pinning landscape, offering an indirect yet experimentally feasible pathway to engineer the coercivity response.

\begin{sloppypar}

\begin{figure*}
  \includegraphics[width=\linewidth]{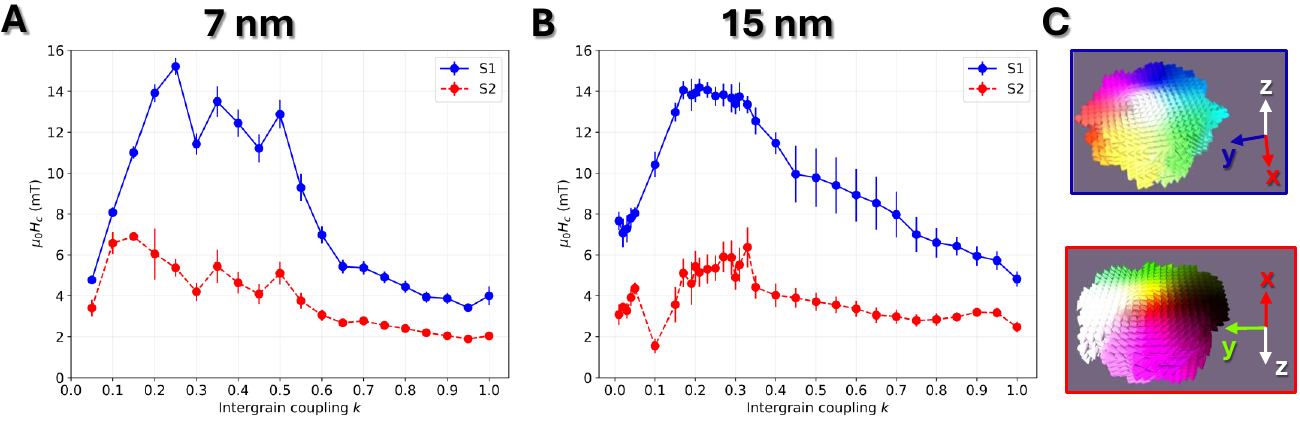}
  \sloppy
  \caption{Influence of inter-grain coupling $k$ on the coercivity evaluated at two different grain sizes in a $d = 100~\mathrm{nm}$ nanoflower. \textbf{(A)} corresponds to 7 nm grain size, \textbf{(B)} to 15 nm. Blue circles (S1) correspond to the formation of a vortex oriented along $z$, while red circles (S2) represent the coercivity for realizations where the vortex core is oriented perpendicular to $z$ at remanence. In \textbf{(C)}, we have included a snapshot of each situation, top corresponding to the core along $z$, bottom, perpendicular to $z$.}
  \label{fig_3}
\end{figure*}

Figures~\ref{fig_3}\textbf{(A)} and \textbf{(B)} show the coercive field versus the inter-grain coupling factor $k$ for grain sizes of 7 and 15 nm, respectively, while the nanoflower diameter is fixed at $d = 100~\mathrm{nm}$, i.e., within the hyperthermia sweet spot. For the 7 nm grains, the inter-grain surface area is roughly twice that of the 15 nm case. In both systems a vortex state exists at remanence, and the coercive field peaks at $\mu_{0}H_{C} \approx14~\mathrm{mT}$ already for relatively weak coupling ($k\approx 0.25$). Because $d = 100~\mathrm{nm}$ lies near the crossover between core-dominated and flux-closure-dominated reversal, some realizations switch through immediate core reversal (vortex core essentially pointing along $z$ direction), while others switch through the second mechanism described in Section~\ref{segunda}, the reversal mechanism with flux-closure rotation (vortex core pointing perpendicular to $z$). In our analysis, we split the data according to the identified reversal mechanism (S1 and S2), visualizing these branches separately in the figure: the higher coercivity branch (blue line S1; upper snapshot in \textbf{(C)}) corresponds to core-driven reversal, whereas the lower branch (red line S2; lower snapshot in \textbf{(C)}) stems from rotational switching.
\end{sloppypar}

\begin{sloppypar}
Note how for these latter flux-dominated realizations (S2, red line), the coercivity remains almost unchanged for both 7 nm and 15 nm grains. This is a consequence of the rotational reversal mechanism, which leaves the vortex core pointing perpendicular to $z$ at remanence, giving only a small net moment and hence coercivity. This kind of reversal dynamics is less affected by grain boundaries, yielding the absence of any significant dependence on $k$.
\end{sloppypar}

Furthermore, the similar coercivity for 7 nm and 15 nm grains in this S2 points to a balance between the number of grains, $N$, and pinning density. In this way, for 7 nm grains, with a larger number of grains, the effective anisotropy is reduced, and so is the coercive field,  $\mu_{0}H_{C}$, scaling as $ \propto K_{\mathrm{eff}} \propto 1/\sqrt{N}$. However, the richer number of grains $N$ implies a larger total grain-boundary area, which grows as $\propto N^{1/3}$, and thus, supplying more pinning sites. The similar (but not identical) weak scaling of these competing effects leaves the coercivity virtually unchanged across the two grain sizes.

\begin{sloppypar}
When NFs are core-dominated (S1, blue line) the coercivity is no longer a constant, yet it describes an arch-shaped pattern with a maximum achieved at intermediate exchange coupling strengths: $k =0.25\text{–}0.50$ for 7 nm grains and $k = 0.15\text{–}0.40$ for 15 nm. This shape is surprisingly universal across several similar yet distinctly different systems. For instance, grain textured nanocomposites show a coercivity maximum set by geometry~\cite{erokhin2017optimization}, and magnetite nanoclusters resembling nanoflowers display a peak at weak coupling followed by a steady drop at strong exchange~\cite{kure2025hysteresis}.

The trend reflects a subtle balance between pinning and coherence of the vortex profile. Starting from the case where the grains are strongly coupled (large $k$), the pinning wells become shallower: the strong exchange reduces both the depth of individual pinning sites and smooths the magnetization, averaging out over multiple pinning sites, further weakening the overall pinning landscape. As a result, the coercivity stays at a low value.

As the coupling decreases, the pinning potentials become deeper, the magnetization does no longer average out over the grains, and as a consequence, the coercivity raises, describing a broad maximum. When the grains are almost fully decoupled (below $k = 0.25$ and $k =0.15$, respectively), however, there is a very low exchange energy penalty for adjacent grains to have a misalignment in their magnetization, resulting in the loss of a rigid vortex profile. In this situation, such low coupling allows the (dipolar field assisted) switching of individual grains, resulting in a significantly low coercivity. We have included results for the extreme case of $k = 0$ in Supplemental Material S6, where the coherence of the vortex profile is completely lost, and instead, a clear multi-domain pattern, in which each grain behaves as an independent single-domain particle with its own random easy axis. In this case, the random direction at each grain cancels out, and a zero coercivity, corresponding to a true multidomain structure, is found. 

We can conclude from our results that the coercivity is maximized at the intermediate $k$ range, where the vortex remains coherent and encounters deep pinning barriers. Larger grains (Fig.\ref{fig_3}\textbf{(B)}), with fewer boundaries, need stronger pinning to reach coercivities comparable to the smaller-grain case (Fig.\ref{fig_3}\textbf{(A)}), and thus, this $k$ range is shifted to lower values. Our results highlight the key role of inter-grain exchange (grain-boundary pinning) in setting $\mu_{0}H_{C}$, and helps to explain the high heating efficiency of NFs in particular, and rich intra-particle grain boundaries nanoparticles in general, compared to their defect-free counterparts, as predicted by A. Lappas \textit{et al.} \cite{lappas2019vacancy}.

\end{sloppypar}

%In summary, our results identify two complementary strategies to tune the formation of a vortex in NFs (i) increase the density of intra-particle boundaries by reducing the grain size, which both lowers $K_{eff}$ and enhances pinning; and (ii) maintain a relatively larger grain size, but reducing the exchange coupling across grain boundaries (i.e., decrease $k$), which maintains the coercivity through boundary pinning.

%Compara unflowered con el trabajo de simulaciones de \cite{concas2024magnetic}, ellos hacen nanoshpheres. Es interesante ver si la forma cambia algo.

%\textcolor{red}{Coercivity goes up when the vortex is set: Let's assume the vortex as a single domain, where the in-plane moments do not contribute to the rotation. What is happening. We have a net magnetization along $z$. When the field rotates it, it does it over Zeeman, so $\propto \mu \times B$. The NF consists of a collection of grains, with random orientations where the core can get pinned. As we increase the NF size, we are increasing the internal boundaries, we are increasing the pinning potentials, the resistance of the core to rotate. This translates into a raise in the coercivity, even if the Vcore/VNF decreases. When the IP moments dominate, the story changes, they are the IP moments which dominate, which carry no net magnetization, and the coercivity drops accordingly. For the case of the perfect sphere, the situation is slightly different in the first regime. The abscence of internal pinning potentials triggers the continuous drop of the coercivity: The core finds no barriers inside the particle preventing his rotation, and indeed, the larger the nanoparticle, the smaller the Vcore/VNF, and therefore, the smaller the coercive field.}

%%%%%%%%%%%%%%%%%%%%%%%%%%%%%%%%%%%%%%%%%%%%%%%%%%%%%%%%%%%
\section{Conclusion}

In this work, we have provided the complete map of the coercivity \textit{vs.} size dependence for one of the most promising candidates for magnetic hyperthermia applications, magnetic nanoflowers. Beyond the single domain Stoner-Wohlfarth regime, we describe the onset of a magnetic vortex, often referred to as multi-domain phase. In nanoflowers this phase comprises a secondary maximum of the coercivity at sizes around 100-150 nm nanoparticle diameter, providing a window for maximized hyperthermia efficiency. 

We examined the reversal mechanism on both sides of the vortex coercivity peak and found two different processes. When the vortex core, which carries the nanoflower’s net magnetization, dominates the particle volume, switching occurs via the immediate core reversal. Once the flux-closure region becomes dominant, reversal proceeds through a smooth rotational process, in which the core aligns perpendicularly to the applied field direction. As the nanoflower diameter increases, the fraction of the volume taken up by the core shrinks, the net magnetic moment falls, and the coercive field decreases, following a power law $d^{-\alpha}$ with $\alpha\approx1.5$–$1.3$; the decay steepens as intra-particle disorder decreases.

We identified the interplay between two features of the grains typical for the nanoflower morphology as the source of the secondary peak in the coercivity. The combination of randomly oriented grain easy axes and weakened inter-grain exchange creates pinning potentials that hinder core reversal, producing a coercivity peak that is absent in non-disordered spherical nanoparticles. We also modeled how changing grain size, an experimentally accessible way to tune inter-grain boundaries, affects coercivity. Halving the grain size doubles the boundary area, boosting the pinning, yet this effect is counteracted by the larger number of grains, thus averaging their random easy-axis orientations and lowering the effective anisotropy. Even though the size of the coercivity peak does not depend on the material disorder, its position does, and is shifted from 150 nm for fully coupled grains to 110 nm for weakly coupled ones. Despite the fact that the exchange coupling cannot be experimentally tuned, engineering the grain size in the nanoflower synthesis is possible and provides a certain degree of tunability to tailor the hyperthermia window, as a larger number of grain boundaries with relatively weak coupling reduction has the same effect on the effective exchange as a lower number with stronger coupling reduction.

\begin{sloppypar}
Our findings clarify why nanoflowers excel as hyperthermia agents. Their unique morphology, rich in pinning sites and local random anisotropy directions, results in favorable stabilization of a vortex beyond the Stoner-Wohlfarth limit, triggering a secondary peak in their coercivity, absent in the typical nanoparticles, but that allows the nanoflowers to perform efficiently within a size range above the single-domain limit. This offeres them greater thermal stability, making them also less prompt to agglomeration, both critical factors influencing their heating performance \cite{wu2024roadmap, serantes2021nanoparticle, etheridge2014accounting}. Furthermore, the specifics of the material disorder, like grain size, allows to finetune the size range of this regime, thereby, optimal heating can be achieved in a broad range of sizes. We anticipate that the insights into the fundamentals of the nanoscale magnetization will guide the design of even more effective nanoflowers for magnetic hyperthermia applications. \newline
\end{sloppypar}

\section{Methods Section}

Micromagnetic simulations were performed using Mumax3 \cite{Vansteenkiste2014}. The nanoflowers were generated from a perfectly spherical geometry with an initial diameter of $d= 10-400~\mathrm{nm}$. A Voronoi tessellation was applied to this sphere to define discrete regions \cite{Lel2014}, corresponding to individual material grains. The grains with Voronoi centers laying inside the base sphere were kept in their entirety (also extending  outside the sphere), whereas areas of grains whose Voronoi center lays outside of the sphere were completely discarded. This results in the non-spherical flower shape as seen in Figure~\ref{fig_1}. We checked that the average volume of the randomly generated geometries coincides with that of the sphere, thus resulting in the same effective diameter.
Each grain was assigned a randomly oriented anisotropy axis to reproduce the polycrystalline nature typically observed in experimentally synthesized nanoflowers \cite{jefremovas2021nanoflowers, moya2024unveiling, bender2018relating, storozhuk2021stable}. 
To reproduce the effect of intra-particle disorder, the magnetic coupling between adjacent grains was modeled by rescaling the exchange parameter $A$ at the grain boundaries by a factor $k$ lying between 0 and 1. We find good agreement with the experimental results for $k = 0.25$ \cite{moya2024unveiling, jefremovas2021nanoflowers, gavilán2017formation, storozhuk2021stable}. Material parameters typical for iron oxide were used based on literature. These include saturation magnetization $M_{s} = 400 \times10^{3}~\mathrm{A/m}$  \cite{roca2007effect, shokrollahi2017review}, exchange stiffness $A=10~\mathrm{pJ/m}$ \cite{sinaga2024neutron}, and uniaxial magnetocrystalline anisotropy $K_{u} = 10^{4}~\mathrm{J/m^{3}}$ \cite{gross2021magnetic, borchers2025magnetic, pisane2017unusual, roca2007effect}. Each simulated size has been realized between 15-25 times with a different random seed for the Voronoi tesselation to ensure the statistical significance of our results. Extended information is included in the Supporting Information S1.\newline

%%%%%%%%%%%%%%%%%%%%%%%%%%%%%%%%%%%%%%%%%%%%%%%%%%%%%%%%%%%%%%%%%%%%%%%%%

\textbf{Supporting Information} \par 
Supporting Information contains extended technical details on the simulations (S1); the full hysteresis loops for selected sizes (S2); the coercivity vs. nanoflower size mapping considering a multi-grain structure for all NF sizes (S3); extended discussion on the determination of the net magnetization direction of the magnetic nanoparticles (S4); the magnetization reversal studied in the case of perfect nanospheres; and the $k =0$ case for the influence of intra-particle disorder.

%%%%%%%%%%%%% Data

\section*{Data availability}

All data supporting the findings of this study are included within the article and any Supporting Information.

%%%%%%%% Conflicts of interest

\section*{Conflicts of interest}

There are no conflicts to declare.\newline

% Acknowledgements

\section*{Acknowledgements}
E.M.J. acknowledges funding from the European Union’s Horizon 2020 research and innovation program under the Marie Sk{\l}odowska-Curie Actions grant agreement 101081455 -- YIA; the Institute for Advanced Studies (IAS) of the University of Luxembourg; and the Fonds Wetenschappelijk Onderzoek (F\-WO-\-Vlaanderen) project number V501325N. 
J. L. is supported by the FWO-Vlaanderen with senior postdoctoral research fellowship No. 12W7622N. 
The computational resources and services used in this work were provided by the VSC (Flemish Supercomputer Center), funded by the Research Foundation Flanders (FWO) and the Flemish Government – department EWI; and from the HPC facilities of the University of Luxembourg and the Luxembourg national supercomputer MeluXina. \newline

%%%%%%%%%%%%%%%  References  %%%%%%%%%%%%%%%%%%%%%%%

\bibliography{references}

\end{document}